%% file: main.tex
\documentclass[conference]{IEEEtran}
\usepackage{ifpdf}
\usepackage{cite}
\ifCLASSINFOpdf
\else
\fi
\usepackage{graphicx}

\newcommand\comment[1]{}

\begin{document}
\title{Towards Understanding Cyberbullying Behavior in a Semi-Anonymous Social Network}

\author{\IEEEauthorblockN{Homa Hosseinmardi,  Richard Han,\\ Qin Lv and Shivakant Mishra}
\IEEEauthorblockA{Department of Computer Science\\
University of Colorado at Boulder\\
\{homa.hosseinmardi,richard.han,qin.lv,shivakaht.mishra\}@colorado.edu}
\and
\IEEEauthorblockN{Amir Ghasemianlangroodi}
\IEEEauthorblockA{Department of Electrical, Computer and\\Energy Engineering\\
University of Colorado at Boulder\\
amir.ghasemianlangroodi@colorado.edu}}


%Telephone: (800) 555--1212, Fax: (888) 555--1212}


\maketitle

\input{abstract}



% \fi
\IEEEpeerreviewmaketitle

\input{intro.tex}

\input{datacllctn.tex}
\input{askfm.tex}
\input{data.tex}
\input{usernet.tex}
\input{wordnet.tex}
\input{netstat.tex}
%\input{ctype.tex}
\input{hrisk.tex}
\input{conclusion.tex}

\bibliographystyle{IEEEtran}
\bibliography{mybiblio}

\end{document}

%% file: abstract.tex
% !TEX root = ICWSM_paper.tex
\begin{abstract}
%\begin{quote}
Cyberbullying has emerged as an important and growing social problem, wherein people use online social networks and mobile phones to bully victims with offensive text, images, audio and video on a 24/7 basis.  This paper studies negative user behavior in the Ask.fm social network, a popular new site that has led to many cases of cyberbullying, some leading to suicidal behavior. We examine the occurrence of negative words in Ask.fm's question+answer profiles along with the social network of ``likes" of questions+answers. \comment{ These are then related to instances of cyberbullying.}  We also examine properties of users with ``cutting" behavior in this social network.

\comment{
One of the most pressing problems in American high schools is bullying. However, with today's technology like cell phones and social networks, bullying is moving beyond the schoolyards to 24/7 cyberbullying in online and mobile social networks, using text, images, audio, and video.  Studying and understanding cyberbullying in social networks can help us to eventually develop automatic detection and prevention tools. Towards these ends, we study in this paper the Ask.fm semi-anonymous question/answer social network, which has gained popularity among teenagers. A number of suicides have been tragically linked to cyberbullying on this website during the past year, underlining the urgency of examining patterns in cyberbullying behavior on this website.  We describe how we have built and used a contextual word group to give us an insight about the language and types of cyberbullying occurring in this website.  We show that an examination of negative words obtained from word graph reveals that there exist two types of victims: first, the group that receives positive posts and support from bystanders; and second, the group without positive questions in their profile pages. Our analysis demonstrates that the second group, which is more vulnerable to cyberbullying, has on average clearly the lowest activity (outdegree) and support (indegree) compared to other defined groups. 
}

%\end{quote}
\end{abstract}

%% file: intro.tex
% !TEX root = main.tex
\section{Introduction}
One of the most pressing problems in high schools is bullying. However, with today's technology, bullying is moving beyond the schoolyards via cell phones, social networks, online video and images, etc. As bad as fighting and bullying were before the prevalence of personal technology, the recording and posting of hurtful content has magnified the harmful reach of bullying. On average, 24\% of high school students have been the victim of cyberbullying \cite{ref1}. Cyberbullying happens in many different ways, including: mean, negative and hurtful comments, pictures or videos posted online or on cell phones, or through the spread of rumors or threats via technology.
 
Although cyberbullying may not cause any physical damage initially, it has potentially devastating psychological effects like depression, low self-esteem, suicide ideation, and even suicide \cite{Hinduja,Menesini}. For example, Phoebe Prince, a 15 year old high school girl, committed suicide after being cyberbullied by negative comments in the Facebook social network \cite{Pheobe}. Hannah Smith, a 14 year old, hanged herself after negative comments were posted on her Ask.fm page, a popular social network among teenagers \cite{hannah}. Cyberbullying is such a serious problem that nine suicides have been linked with cyberbullying on the Ask.fm Web site alone \cite{askfmsuicides}. \comment{Although cyberbullying is not the direct cause of these suicides, }Cyberbullying was viewed as a contributing factor in the death of these teenagers \cite{ref1}. Given the gravity of the problem and its rapid spread among middle and high school students, there is an immediate and pressing need for research to understand how cyberbullying occurs today, so that techniques can be rapidly developed to accurately detect, prevent, and mitigate cyberbullying.
 
 % leading to efficient detection of potential cybersecurity incidents in near real time. % According to ``A thin line'' program, teenagers believe bullying is  much more horrible cyber-wise, because every one can see it and there is nothing you can do to hide it. Not only are cyber attacks more public, they are often more vicious when the attacker is not face to face with the victim. It become worse when it comes anonymous. For example, Adalia Rose, a 5-year old girl with a rare disease, has been bullied on her Facebook page and hate pages have been created for her \cite{Adalia}. A BBC documentary in 2013 states that the rate of cyberbullying is increasing. The same documentary reports that 28\% of 11-16 year old teenagers have experienced bullying or harassment online \cite{BBC}
 While most current studies have focused on the prevalence and impact of cyberbullying in education and psychology \cite{4,5,6}, our interest is in understanding how social networks are being used to enable cyberbullying.  
\comment{ are interested to first analyze online social networking data to understand
the key distinguishing features of cyberbullying posts in an online social network. 
We should keep it in mind that there is no single feature that can accurately identify a cyberbullying incident, and we will have to rely on a combination of several different features, perhaps each with a different weight. 
}
Prior work in cyberbullying analysis and detection in social networks has largely focused on such social networks as Youtube, Formspring, MySpace, and Twitter~\cite{Dinakar_modelingthe2, dadvar, Twitter}.     ~\cite{Dinakar_modelingthe2} investigated both explicit and implicit cyberbullying by analyzing negative text comments on Youtube and Formspring profiles.  ~\cite{dadvar} investigates how integration of MySpace user profile information like gender in addition to text analysis can improve the accuracy of cyberbullying detection in social networks   ~\cite{Twitter} tries to detect bullying in Twitter text data by looking for inappropriate words using a Naive Bayes classifier. They track potential bullies, their followers and the victims. All of these works focused on text-based analysis of negative words, and did not exploit social network relationships in their investigation of cyberbullying.

\comment{
 A key property of these networks is that social graphs can be built by studying the friendship connections between users.  In contrast, a network such as Ask.fm is an example of a semi-anonymous social network.  In such a network, posters by default post comments anonymously on another person's profile.  As a result, it is difficult to build a social graph based on friendships, because these are not clearly expressed.  Our challenge 

just look at text, whereas we look at the graph

ADD EXAMPLES of CYBERBULLYING ANALYSIS IN FACEBOOK, TWITTER, ETC.

NEXT, SHOW WHY WHAT WE ARE DOING IS DIFFERENT

SHOW WHY THIS IS HARD IN ASK.FM, a SEMI-ANONYMOUS SOCIAL NETWORK
}
Previous work \cite{myspace} has considered characterizing the language model of MySpace users, though the emphasis was not on negative word usage. Our work seeks to understand whether social network relationship information can be useful in supplementing purely text-based analysis in helping to identify negative user behavior in social networks. We are interested in building graphs from social networks and analyzing their properties to extract features such as in-degree or out-degree that may be useful in flagging such bad behavior.

\begin{figure*}[!ht]
\begin{minipage}[b]{0.48\linewidth}
\centering
\includegraphics[width=\textwidth]{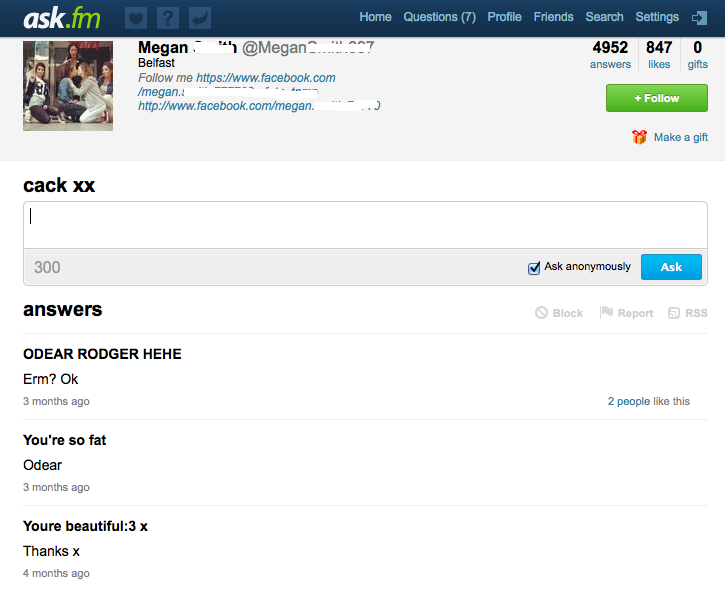}
\caption{Typical public profile on Ask.fm with questions/comments (usually anonymous) to the profile owner and answers from the profile owner.}
\label{figintro}
\end{minipage}
%\hspace{0.5cm}
\hfill
\begin{minipage}[b]{0.48\linewidth}
\centering
\includegraphics[width=\textwidth]{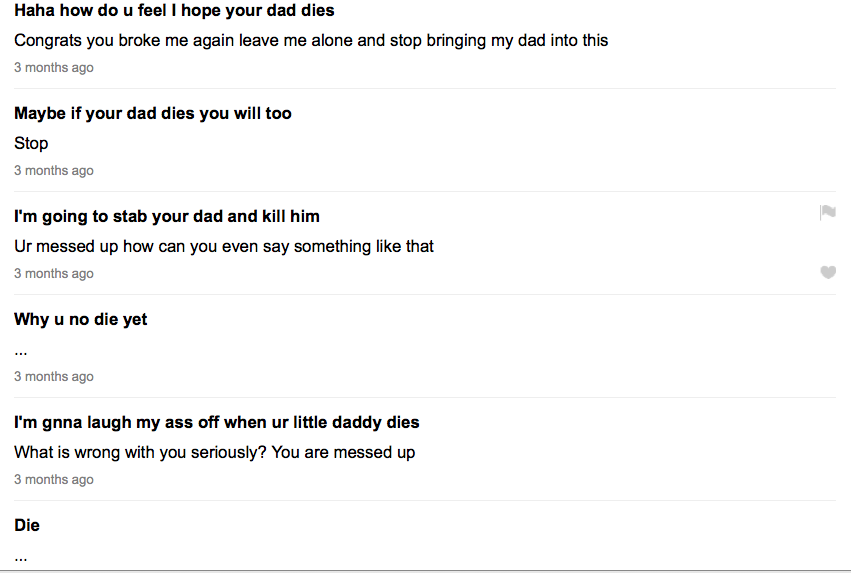}
\caption{An example of anonymous cyberbullying comments posted on a user's profile in Ask.fm.}
\label{figrepeated}
\end{minipage}
\end{figure*}

In particular, this paper chooses to focus on analyzing the Ask.fm social network for the following key reasons.  First, Ask.fm is a major source of cyberbullying on the Internet.  In fact, it ranks as the fourth worst site in terms of percentage of young users bullied according to a recent survey~\cite{annual}, after Facebook, YouTube and Twitter.  Second, very little research has been published to date concerning the Ask.fm  social network.  Third, Ask.fm is a highly popular and rapidly growing social network,  with over 70 million registered users as of August 2013 \cite{askfmwiki}.  Finally, Ask.fm provides publicly accessible data.

One of our big challenges in analyzing the Ask.fm social network is that it behaves as a semi-anonymous social network.  User profiles are public, but postings to each profile by users other than the owner are by default anonymous.  In addition, we cannot obtain from the public profiles which users are following which other users.  \comment{When collect info on profile, \# of people follow is not known.  Just profile owner knows how many are following him.  But you're not logged in.  Ask.fm you can follow people, but this is anonymous.}  As a result, it is not possible for us to construct a social graph based on friendships.  However, we observe that another type of graph called an interaction graph \cite{Wilson} can be extracted from the ``likes" of comments.  We use this insight to build and analyze interaction graphs that embed social relationship information, to help identify negative user behavior in this semi-anonymous social network.

%Interaction graph... not building a social graph... two types of graphs...

This paper makes the following contributions.  It is the first paper to to provide a detailed characterization of key properties of the important Ask.fm semi-anonymous social network.  Second, it builds and analyzes interaction and word graphs and finds that properties of the interaction graph such as in-degree and out-degree are strongly related to the amount of negative user behavior expressed on a profile, i.e. highly positive profiles exhibit the highest degree of sociability in terms of liking others and being liked by others, whereas profiles with a high number of negative questions exhibit the lowest degree of sociability.

In the following, we describe our data collection efforts, build graphs of users' ``likes" as well as negative word graphs, and use these to illustrate the relationship between negative words and user activity.  We also analyze a particularly high risk set of users who state on their profile that they have ``cut" themselves.  \comment{We also use these graphs to illustrate the differences between various types of cyberbullying, e.g. racial, sexual, religious, appearance-based, etc.}
% We believe this site will only become increasingly popular among teens - a population especially prone to and victims of cyberbullying - since a recent trend has been that teens are moving away from well-known social networking sites like Facebook that their parents know about and towards other social networks that are not as well-known to their parents [16], [17], [18], [15], such as Ask.fm.

%% file: datacllctn.tex
% !TEX root = main.tex
\section{Data Collection}

In this section, we discuss key aspects of the Ask.fm social network,  what information was collected from Ask.fm's publicly accessible profiles, e.g. questions, answers, and likes, and our process of building interaction and word graphs.  \comment{how we build a graph when the network is semi-anonymous, namely the identities of posters who comment on an individual's posters are by default unknown.}
% In a typical network, 
%The Ask.fm site has the virtue that its information- questions, answers, likes, etc. - is publicly accessible.  For this study, we sampled the set of 70 million users.  \comment{, which we cannot collect all the user's profiles and the need for sampling arise here. There are several different methods in sampling, such as uniform random node selection, Breadth-First search, Random Walk, Depth-First search, and Forest Fire \cite{mgjoka_recommendations}.} Each of the sampling methods has its own benefits and drawbacks. In fact, choosing a sampling method depends on what interesting properties you want to extract from the dataset. A good sampling is one that does not change the statistics or information of the network that you are interested in. In this paper, we want large enough samples to be able to extract negative and non-negative behavior patterns. 

%% file: askfm.tex
% !TEX root = main.tex
\subsection{Ask.fm}
We observe that Ask.fm is generally an example of a semi-anonymous social network, in that the
identity of users who post questions/comments to a profile is typically anonymous (though posters
may choose to reveal their identity, we have seen this happen only rarely), whereas the identity of
the target user is publicly known. People can search Ask.fm users via their name, id or email address. This is unlike non-anonymous social networks such as Facebook
and Twitter, where the identity or ID of posters and target users are both publicly visible. Thus, in
Ask.fm, any one (even people without an account) may post on another user's profile, and this posting of a question or comment is
usually done anonymously (in fact, that is the default). 

There are some policies specific to Ask.fm.  First, only the target user may post an answer to a question/comment on this site.  Only after answering a question will the question and answer appear on his/her profile. Further, a user may choose to ``like'' at the granularity of a question+answer pair, but cannot like the question \comment{individually}nor the answer individually.  Liking is non-anonymous, so that
the identity of the likers is publicly known. Another feature of Ask.fm is that users may follow other users.  \comment{, and in this case posted questions+answer of those who are following will posted on the user's home page.}  However, this relationship data is not available publicly and only the profile owner knows who he/she is following. Even the user who is being followed can only know how many followers he/she has (which is not publicly available to other users), not who is following him/her. \comment{In this case by crawling publicly available data, we can not claim our dataset has complete information about a profile.}

Figure \ref{figintro} shows a typical publicly accessible profile obtained from the Ask.fm social Web site.   We see that other users may post questions/comments on a target user's profile, and that the target user may answer each question/comment.  In this example, we see both a negative comment ``You're so fat" mixed in with a positive comment ``You're beautiful".  A more serious example of cyberbullying from the Ask.fm Web site is shown in Figure \ref{figrepeated},  where we observe that anonymous negative comments have been repeatedly posted on the target's profile wishing or threatening death upon the target's father.
%How people socialize in Ask.fm? In this social network users are not connected by friendship. There is ``FOLLOW'' option which most of the users; especially the ones that we are interested in (victims receiving bullying posts) do not have usually followers. There are also two concepts for interaction of the users: 1- question, 2- like. Questions are posted mostly anonymously, however there exists the option of non-anonymity in posting question. The only ways to socialize with other users non-anonymously in this social network is the ``like'' button. For each question and its following answer by the user, there is only one ``like'' button which adds another level of ambiguity to this data. 

%% file: data.tex
% !TEX root = main.tex
\subsection{Description of Collected Data}

The data we can extract from a common profile includes the following fields: userID, personal information (if any, as it is optional), total number of answers, total number of likes, content of answered questions posted on a user's page, and the userID of people who liked the questions+answers.

An interaction graph can be constructed from the ``likes" of answered questions, i.e. each directed edge in the graph connects user $i$ to neighbor $j$ if user $i$ has liked a question+answer pair in $j$'s profile.  Note that the edges are not bidirectional, so that $i$ liking an question+answer on $j$'s profile does not imply $j$ liking one of $i$'s question+answer.  In order to extract this interaction graph from Ask.fm, we conducted a breadth-first search starting from a couple of random seed nodes. Seed nodes should have non-zero liked question+answer pairs. For each seed node we found all nodes that liked an answered question on its page (incoming edges are publicly known for each profile). However by only looking at a user's profile, we are not privy to any of the nodes that this profile owner liked (outgoing edges).  In the second step, we collect the profile information of all neighbors of the seed node.  \comment{new observed but not-sampled nodes. }These steps are then repeated.  Ultimately, the outgoing edges from a profile can be reconstructed from the incoming edges of other profiles.   Note that because we're crawling profiles using breadth-first search, we can only find that subset of the outgoing edges for each profile that happen to be incoming edges on other crawled profiles.  The only way to find all outgoing edges for each profile is to crawl the entire Web site, which is impractical.  \comment{Profiles with zero liked question+answer pairs will not be collected in this approach.} \comment{ If this method does not repeat enough to cover the entire network, then we are looking densely at some specific part of the graph within a distance from the starting point \cite{mgjoka_recommendations}.  At the end we have all the links between sampled nodes, and just incoming links to observed nodes.}  Since the breadth-first search is terminated before crawling all profiles, then this is called snowball sampling, and results in an interaction graph wherein all the internal nodes have been fully crawled or sampled, but also a typically small fraction of nodes on the edge that have not yet been crawled.  Our analysis below focuses only on fully sampled nodes in the interaction graph.  Using snowball sampling, 30K profiles were crawled from October to December 2013. \comment{which is about 0.2\% of the available profiles on Ask.fm.  Other studies on social networks have sampled about 0.3\% of Orkut users and 0.08\% of MySpace users \cite{Ahn}.}

\comment{Here, unlike a typical non-anonymous social network, we lack friendship information in a semi-anonymous social network. People a user follows them, known as his/her friend is only visible to the user. Fortunately, we note that liking others' comments and having others like comments on your profile is non-anonymous in Ask.fm. Therefore, it is possible to construct an interaction graph based on likes, where the nodes in the graph are the users, and the edges between nodes are directional, showing that one user has ``liked" a question or question+answer pair of another user. In this case we can build the graph based on users who actually have interaction which each other. The liking relation is a directed behavior, therefore our graph will be directional. Author in \cite{Wilson} consider two types of graph for Facebook social network. One named social graph, build based on friendship, and the other named interaction graph, connects two nodes if any of them post on the other one's wall or photos. }

\begin{figure}[htbp!]\centering
\includegraphics[width=0.45\textwidth]{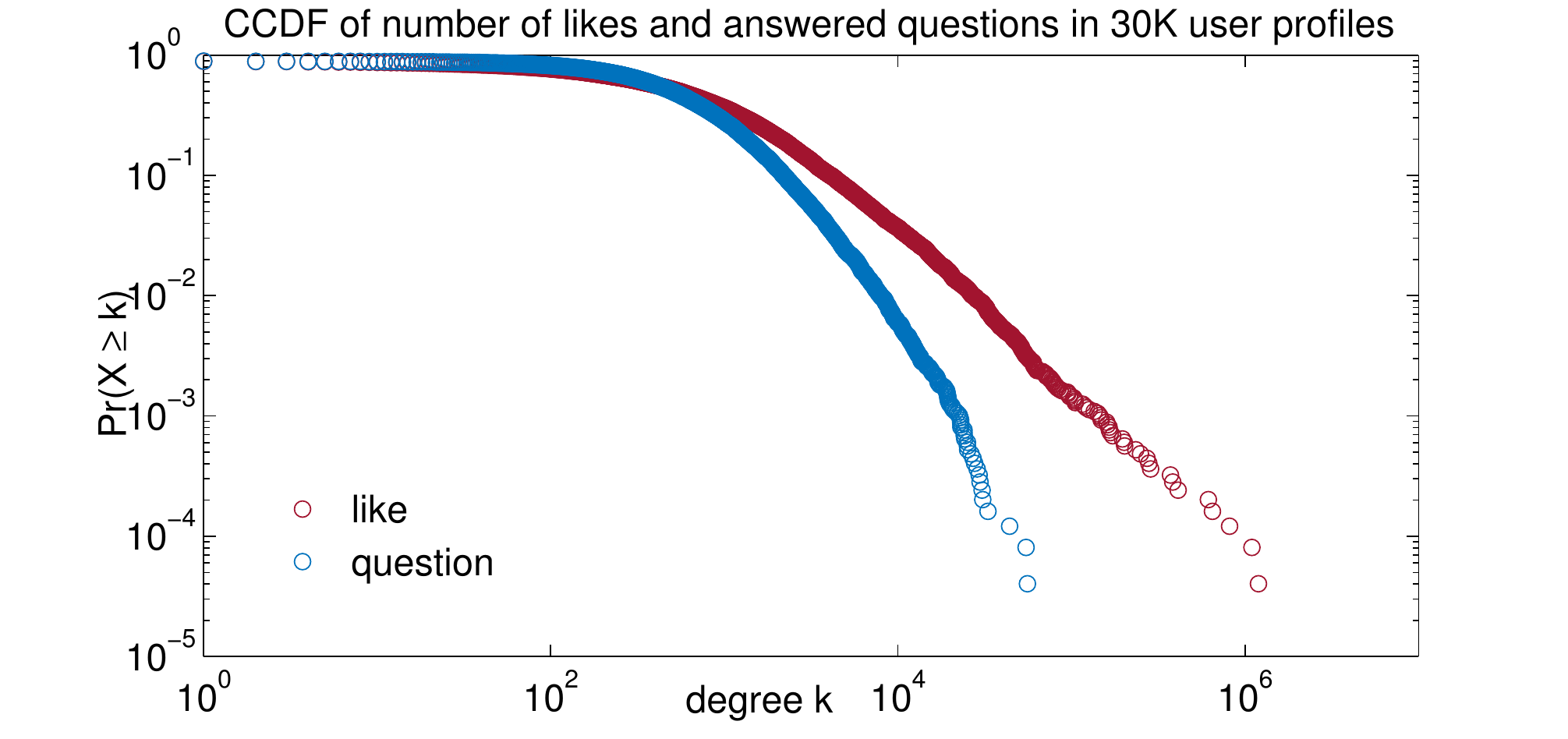}  % use this if you use "pdflatex"
\caption{Probability that the total number of likes and question+answer pairs for a user is greater than or equal to k}
\label{fig1} % Fig.3
\end{figure}

To provide some general context, we first perform an analysis of the total number of question+answer pairs and likes per Ask.fm user.  We observe they both have heavy tail distributions in Figure \ref{fig1}. The total number of likes has a heavier tail than the total number of question+answer pairs. \comment{We can consider two reasons for this observation. When the number of questions is small, the user is either not active, or not popular and receives few questions. In this case the number of likes that these users receive is typically less than 1 per question.} Looking at the tail, we observe that as the total number of answered questions in a user profile increases, the total number of likes also grows, Figure \ref{fig22}. The red line is the best fit obtained through linear regression in the transformed (log-log) space. It seems when a user is more popular and active, they receive more questions, and more user visits and likes on his/her page. Also pushing a like button is easier than writing a question, which makes the number of likes larger than the number of questions.

\comment{
\begin{figure}[htbp!]\centering
\includegraphics[width=0.45\textwidth]{./figure/qstn_like_150K-eps-converted-to.pdf}  % use this if you use "pdflatex"
\caption{Mean number of likes versus total number of answered questions a user receives}
\label{fig2} % Fig.3
\end{figure} 
}

\begin{figure}[htbp!]\centering
\includegraphics[width=0.45\textwidth]{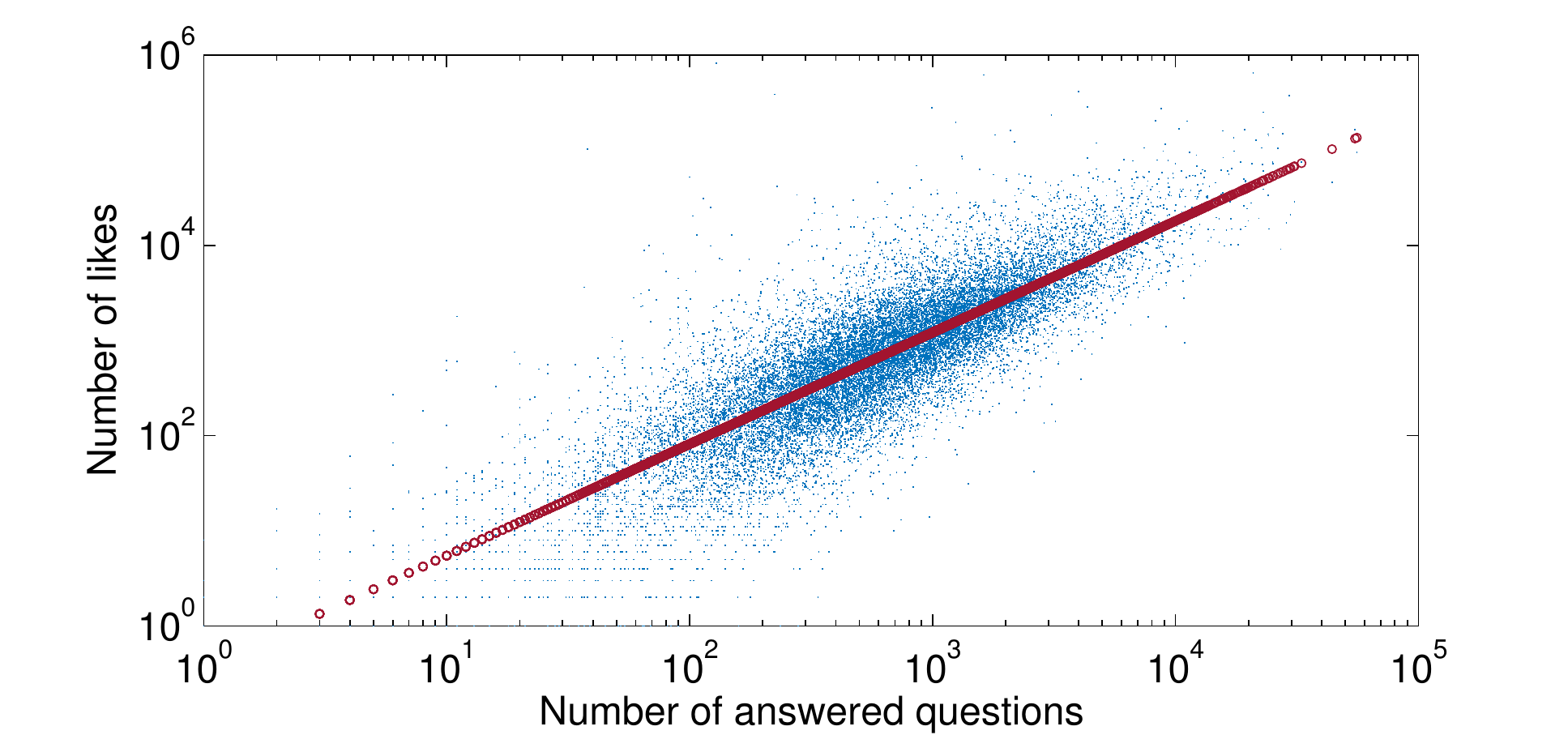}  % use this if you use "pdflatex"
\caption{Total number of likes versus total number of answered questions a user receives}
\label{fig22} % Fig.3
\end{figure}

Figure \ref{fig22} describes the correlation between the number of answered question+answer pairs and the number of likes.  The correlation value when the number of answered questions is less than 50 is -0.05. This shows that when the number of answered questions in a user page is low, the number of people who like these users does not have any special relation with the number of answered questions. When the number of answered questions is larger than 50, the correlation between the number of likes and the number of answered questions increases to 0.33. This means that as the number of question+answer pairs increases beyond around 50, the number of likes becomes weakly related to the number of answered questions. \comment{This interesting phenomenon suggests an acceleration of popularity, wherein as a user receives more questions, their average number of likes per question increases.}

%% file: usernet.tex
% !TEX root = main.tex
\subsection{Modeled Interaction Network}
%We are interested in studying the relationship between properties of the interaction graph and negative user behavior, as revealed by the usage of negative words.  Cyberbullying occurs when there are repetitive negative words and abusive text on profiles.
The preceding analysis provided a general overview of likes and answered questions in Ask.fm, but we are more interested in likes and answered questions in the context of cyberbullying.  Our search for cyberbullying on the Ask.fm is based on the insight that repetitive negative words represent the core of the abusive text posted on profiles. Following the occurrence of negative words led us to many examples of cyberbullying. However, after a preliminary analysis of the answers by the profile owner, we found that many such examples seemed to have no effect on the targeted user, namely based on the answers the target seemed indifferent to the negative comments. In contrast, there were other more serious cases where the target seemed particularly vulnerable, making statements like ``you broke me" (See Figure \ref{figrepeated}).  \comment{Since our greatest concern is in identifying high risk victims of cyberbullying, namely those victims who were most at risk of suicidal behavior, we were most interested in analyzing patterns related to the profiles of the most vulnerable victims of cyberbullying.}

Looking at the profiles, we observed that the users who seemed most vulnerable to negative questions were often those who were most isolated, with few ``likes" and also rarely liking others' comments. In contrast, users who were subject to but seemingly indifferent to negative questions appear to have a fair number of likes and also seemed somewhat active in liking others' comments.

Based on these two observations that (i) cyberbullying is the behavior of posting questions with negative words and (ii) vulnerable targets of cyberbullying (based on their answers) seem isolated, we sought to build and analyze social graphs and word graphs derived from our data that would capture the negativity and isolation of users.
\comment{Looking at the literature, two types of graph can be build for a social network \cite{Wilson}. One named social graph, build based on friendship, and the other named interaction graph, connects two nodes if any of them post on the other one's wall or photos. As we explained in previous section there in no friendship/following data available and question mostly are posted anonymously. The only non-anonymous (non-optional) interaction between users is the act of liking each others question+answer. Therefore, it is possible to construct an ``interaction graph" based on likes, where the nodes in the graph are the users, and the edges between nodes are directional, showing that one user has ``liked" a question or question+answer pair of another user. In this case we can build the graph based on users who actually have interaction which each other. The liking relation is a directed behavior, therefore our graph will be directional.  Due to practical limitation we are not able to crawl all the contents of a profile from the date is has been created and also there are many posts which have not been liked even once. }

\comment{ partial picture: not all outgoing links, only top 15 likes, can't find followers nor who you're following, ...}

\comment{connect a user to a word if that word has been seen in the user's profile...

C. relating likes and answered questions graph structure to negative words

D. is which negative words (and positive)}

In order to capture the greatest degree of interaction between the users, we collected the top 15 questions for each user that had the highest number of likes. For some users with very low activity, the number of highest liked questions was less than 15. Analyzing the most popular questions of profiles, we have built our network modeled as a directed bipartite network. In order to build our graph, if user $i$  likes a question+answer in the page of user $j$, then there is a link from $i$ to words on that question and a link from those words to node $j$. 

From this bipartite graph, we seek to derive characteristics that can highlight the negativity and isolation associated with targeted users.  In order to project a bipartite network with adjacency matrix $B$, to the network of words $W$, we have $W = B B^T$. Then we can similarly build the network of users with adjacency matrix $U$ from our bipartite network. That is, based on this data, our idea is to build and use a like-based interaction graph between users ($U$) and examine the balance of in-degree vs out-degree of the users with high degree of negativity in their pages and small number of positive posts; such users have both low in-degree and low out-degree in graph terminology. In contrast, users who received on average the same amount of negativity but have positive questions at the same time, show healthier in-degree and out-degree. \comment{Thus, we feel that measuring and comparing the in-degree and out-degree of nodes would help us identify and distinguish the most vulnerable victims of cyberbullying.}

%% file: wordnet.tex
% !TEX root = main.tex
\subsection{Extracting Positive and Negative Words}
We next consider which negative words are pertinent to our analysis of negative user behavior.  The natural approach is to select those negative words that have the highest frequency in the sampled profiles.  However, we observe that it is the collective effect of negative words that is exploited by cyberbulliers.  Negative words that may be commonly used but more in isolation rather than in a collective fashion would be less likely to create a strong effect of cyberbullying.  Therefore, we sought to create a word graph that measures the relationship between words, i.e. are they being used together on the same profile to bully a victim.
\comment{Prior work has typically analyzed the top N negative words that occur with the highest frequency ~\cite{MySpace}.

eigenvector centrality measures the connectedness of a node to others nodes in a graph

Collecting a subset of the most important negative words.  To make the analysis more tractable.  One day of continuous compute time for 30 K.
}
\comment{In addition, we seek to derive from our data a word graph ($W$) that shows which words (negative, positive, other) are used on which user profiles.  This will allow us to characterize the negativity associated with each user.  }By one mode projection from a bipartite graph comprised of 30K users and a dictionary of around 1500 negative words (obtained from \cite{NegativeWordsList}), we constructed a word graph wherein each word signifies a node in the graph and each edge indicates that the two words have been used in the same profile.

\comment{
highest in degree in word graph

Instead of looking at the frequency of negative words, we look at pairs of words that have been seen together mostly.  Why?  When you're talking about negativity and cyberbullying, negative words that have been seen together have more impact.  repetition of negative words = cyberbullying, 

only 10 used in MySpace analysis...

correlation lower if based on frequency only...  word "dumb" used in isolation, ... repetitive used in cyberbullying....  highest eig centr captures goal better for cyberbullying than highest frequency...
}

\comment{Figure \ref{figW} illustrates a word graph constructed from these negative words found on Ask.fm user profiles.}  We found that there is a cluster of connected negative words in the center of the word graph, but there are also many negative words that were not connected to any word.  Out of 1500 words, the eigenvector centrality of 968 words is zero, which means 968 negative words either have not been seen in any profile or have not been seen together in a profile. These words were eliminated from our analysis.  \comment{Further, among the remaining words, some were used more frequently than others.}  The top row of table \ref{tbl1} shows the remaining negative words that had the highest frequency of appearance.  \comment{However, focusing only on frequency does not capture the cyberbullying that we witnessed, in which negative words are used repeatedly on the same profile. To capture this effect, we computed the eigenvector centrality, which measures which words are found with other words on the same profile.}  The second row of Table \ref{tbl1} shows the negative words with the highest eigenvector centrality.  \comment{These are high frequency words that are connected to other high frequency words.  }   
Though there is a fair degree of overlap with the highest frequency words, we note that the sets do indeed differ.   Since eigenvector centrality captures to some extent the collective negativity of cyberbullying, we focused our word graph analysis below on the 80 negative words with eigenvector centrality values larger than 0.5. A similar approach was followed for a collection of 1000 positive words.  80 positive words with highest eigenvector centrality were chosen for the following analysis.

\begin{table*}[htbp]\centering
\begin{tabular}{lclclclclclclclclclclclclclclclclclclc|c|} 
hoe & n**ga & stupid & slut & kill & p**sy & cut & suck & gay & fat & sex & bad & d**k & die & ugly & s**t  \\ \hline
f**k & ass & bitch & s**t & hate & d**k & ugly & sex & bad & suck & p**sy & fat & gay & stupid & die & kill \\ \hline
\end{tabular}
\caption{The first row shows the negative words with the highest frequency and the second row shows the negative words with the highest eigenvector centrality (values decrease from left to right).}
\label{tbl1}
\end{table*}

%% file: netstat.tex
 % !TEX root = main.tex
\section{Network Statistics}

\subsection{Building the Interaction Graph}
\comment{As there is no friendship which connect to user, in fact two users are connected if there is an interaction between them. The only non-anonymous interaction among users is liking the question+answer in each other pages.}  
\comment{The preceding analysis of 150K profiles was wide but limited in that  only a limited number of likes were retrieved per question, not the complete list of users who liked a question+answer.  To obtain a deeper and more complete picture of the liking behavior, }
We collected and analyzed  about 30K profiles that had a complete list of likes, using the snowball sampling method\comment{ starting from a random seed node}, \comment{Note that in this section we want to extract patterns of different types of users. Therefore, with this subset of samples the produced bias will affect in-degree, out-degree and etc. of all considered types at the same time. As mentioned, we focused on}gathering the top 15 most liked questions. We found that there were on average 14.5 most liked questions per user, that is most users had close to 15 questions with likes. We also focused on a collection of 80 negative words and 80 positive words with the highest eigenvector centrality as explained previously.

\begin{table}[htbp!]
\centering
\begin{tabular}{|c|c|}
\hline 
Average number of answers per user  &  	 14.5	\\ \hline
Average number of negative questions per user 	  & 0.778\\ \hline
Average number of positive questions per user   &    4.57\\ \hline
Average number of negative words per user 	  & 1.04\\ \hline
Average number of positive words per user   & 5.42\\ \hline
\end{tabular}
\caption{Average number of questions and words per user}
\label{stats}
\end{table}

We found that \comment{Out of 27910 profiles, 12845 users or}46.0\% have at least one question including one of the top negative words. We term such a question a ``negative" question.  Furthermore, \comment{2159 users or }7.74\% of the users have at least three questions with top negative words, and \comment{ 26779 or }96.0\% users had at least one question with a top positive word in their profiles. We term such a question a ``positive" question.  Table ~\ref{stats} shows the average number of positive and negative questions and words per user.

By a one mode projection of the bipartite graph, we were able to obtain the like-based interaction graph $U$ between these 30K users in the Ask.fm social network. \comment{We restricted our analysis to consider likes for only the top 15 most liked questions/answers on each user profile.} In the adjacency matrix $U$, created from the users' interaction graph, edges are weighted. In fact the weight of each edge $e_{ij}$ , from node $i$ to node $j$ is a vector $[n_1 $ $n_2]$, where $n_1$ is the number of questions in node $j$ including at least one negative word (from our selected dictionary) and $n_2$ is the number of questions with no negative word liked by node $i$. Since we are most interested in negative behavior, we group positive and neutral behavior together into a ``non-negative" category. 

\begin{figure}[htbp!]\centering
\includegraphics[width=0.4\textwidth]{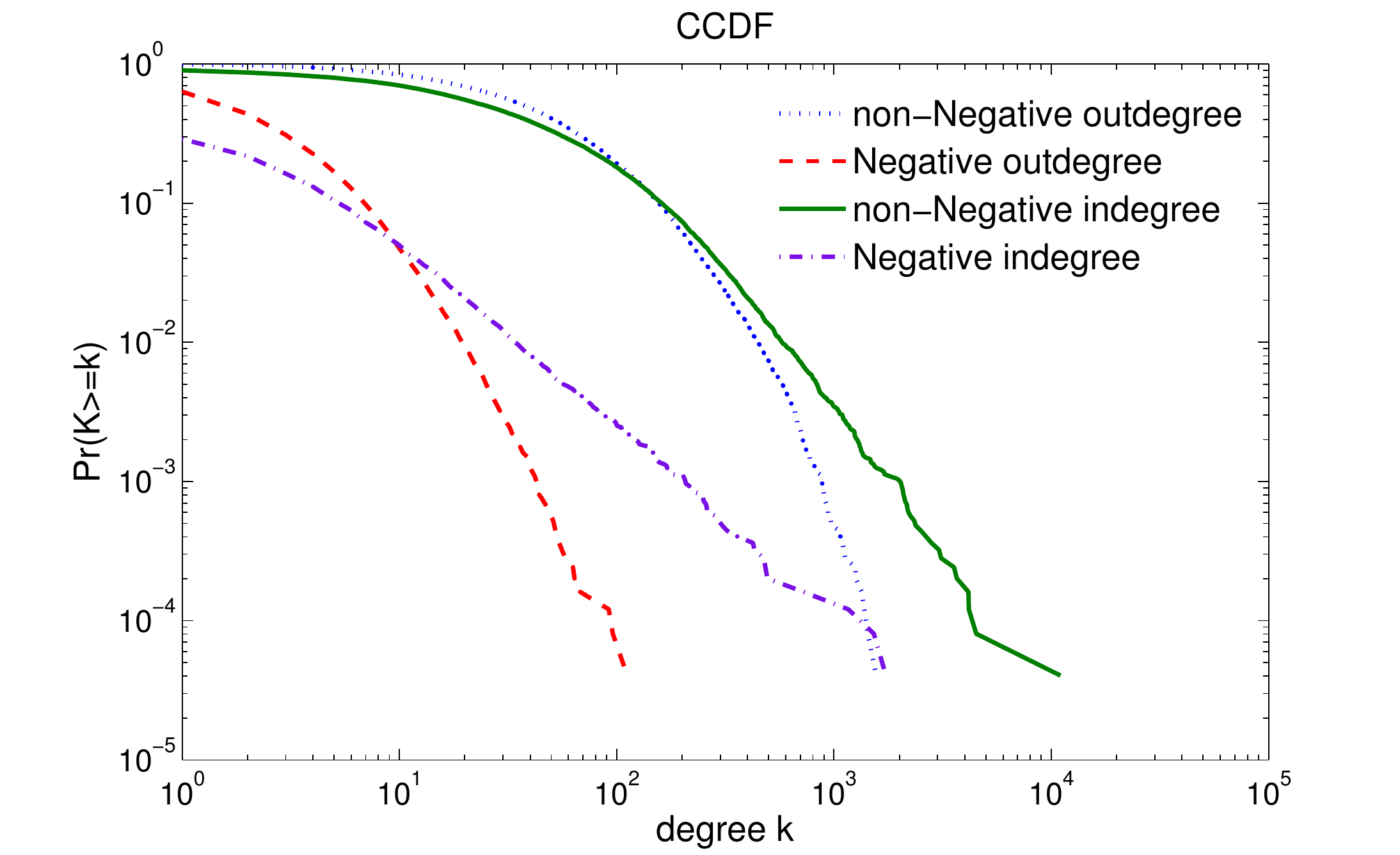}  % use this if you use "pdflatex"
\caption{Complementary cumulative distribution function for negative and non-negative interaction graphs/matrices.}
\label{figDist} % Fig.3
\end{figure}

For further analysis, we have divided matrix $U$ into two matrices $U_{neg}$ and $U_{non-neg}$, where $U_{neg}$  (negative adjacency matrix) is the adjacency matrix with weights $n_1$ and $U_{non-neg}$ (non-negative adjacency matrix) is the adjacency matrix with weights $n_2$. 
% I should double check this paragraph
Figure \ref{figDist} shows the CCDF for the in-degree and out-degree distribution of matrices $U_{neg}$ and $U_{non-neg}$. Authors in \cite{Mislove} show that in-degree and out-degree distributions in social networks are approximately  the same. We can see from Figure \ref{figDist} the in-degree and out-degree for non-negative degree distribution is approximately the same for non-negative degree distributions.   \emph{However, the negative in-degree and negative out-degree distributions of the interaction graphs clearly differ, indicating that negative behavior in the Ask.fm semi-anonymous social network has different properties than previously reported in other social networks}.\comment{ I may add D metric here }

\begin{figure}[htbp!]\centering
\includegraphics[width=0.45\textwidth]{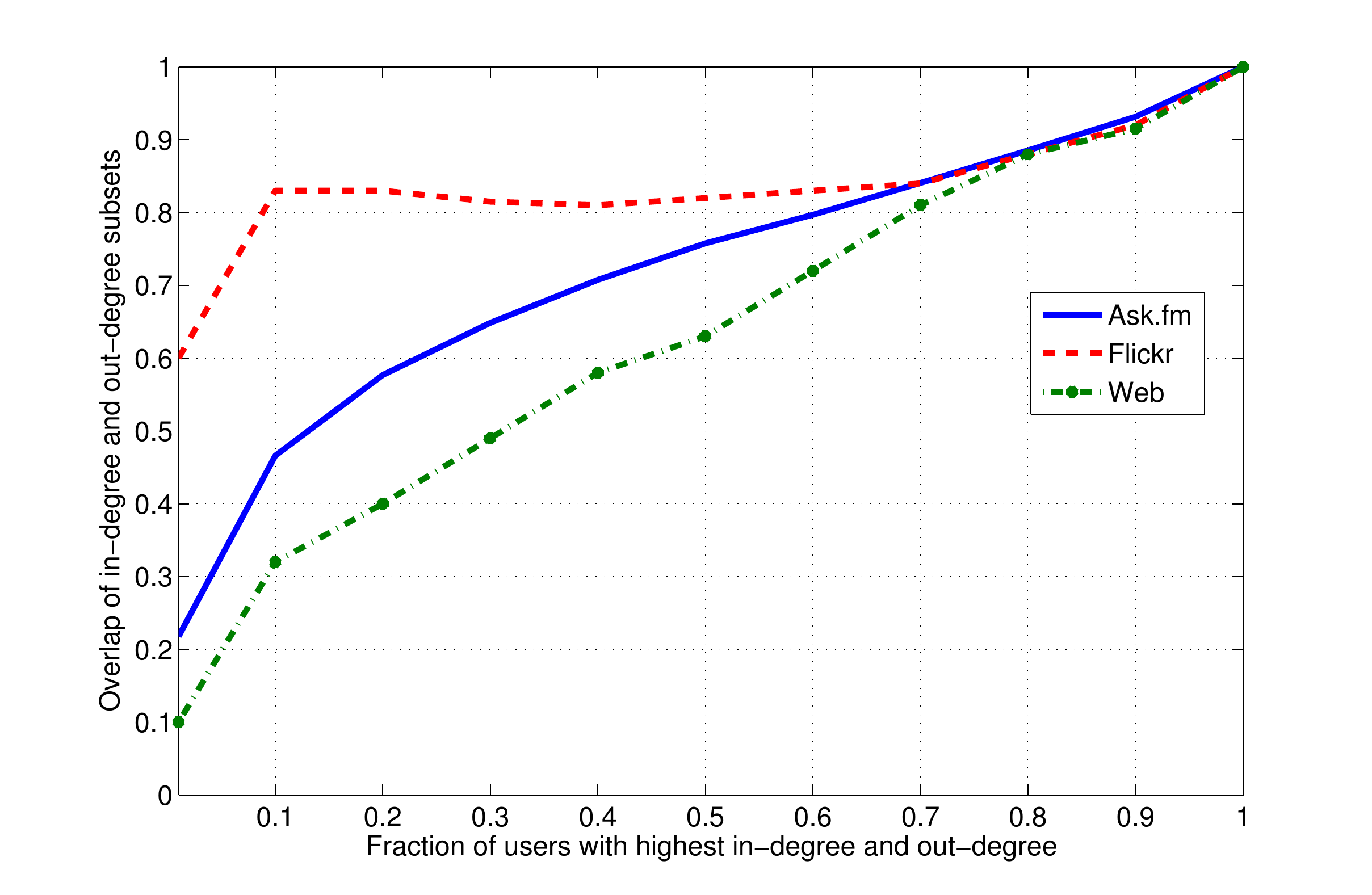}  % -eps-converted-to.pdf use this if you use "pdflatex"
\caption{Percentage of common users among two subsets of x\% highest in-degree and x\% highest out-degree. Measurements for Flicker and Web have been obtained from \cite{Mislove}}
\label{figOverlap} % Fig.3
\end{figure} 

Figure \ref{figOverlap} shows the percentage of common users among x\% highest in-degree and x\% highest out-degree. In fact this figure shows whether the users with high out-degree are the same people with high in-degree. Authors in \cite{Mislove} show in social networks for users with 1\% highest in-degree and out-degree, more than 60\% of users are common.  This value is less than 20\% for the Web. Here we observe that Ask.fm's semi-anonymous social network exhibits behavior that is between previously analyzed social networks and the Web. Overlap of the users with 1\% highest in-degree and out-degree is more than 20\% and less than 30\%. It seems the correlation between in-degree and out-degree in an interaction graph built from like-based interactions is much less than the correlation in friendship-based social graphs. Friendship is usually a symmetric relation. However, in like-based interactions symmetry is less probable, which causes less correlation between in-degree and out-degree. \comment{On the other hand we are dealing with a semi-anonymous social networks and even in the non-anonymous interactions (liking)}In addition, there is the ambiguity in Ask.fm as to whether a like is for a question or answer. This decreases the correlation between in-degree and out-degree because we do not know whether the liker is liking the question as a support of the question's content or the answer to support the profile owner. 

\begin{figure}[htbp!]\centering
\includegraphics[width=0.45\textwidth]{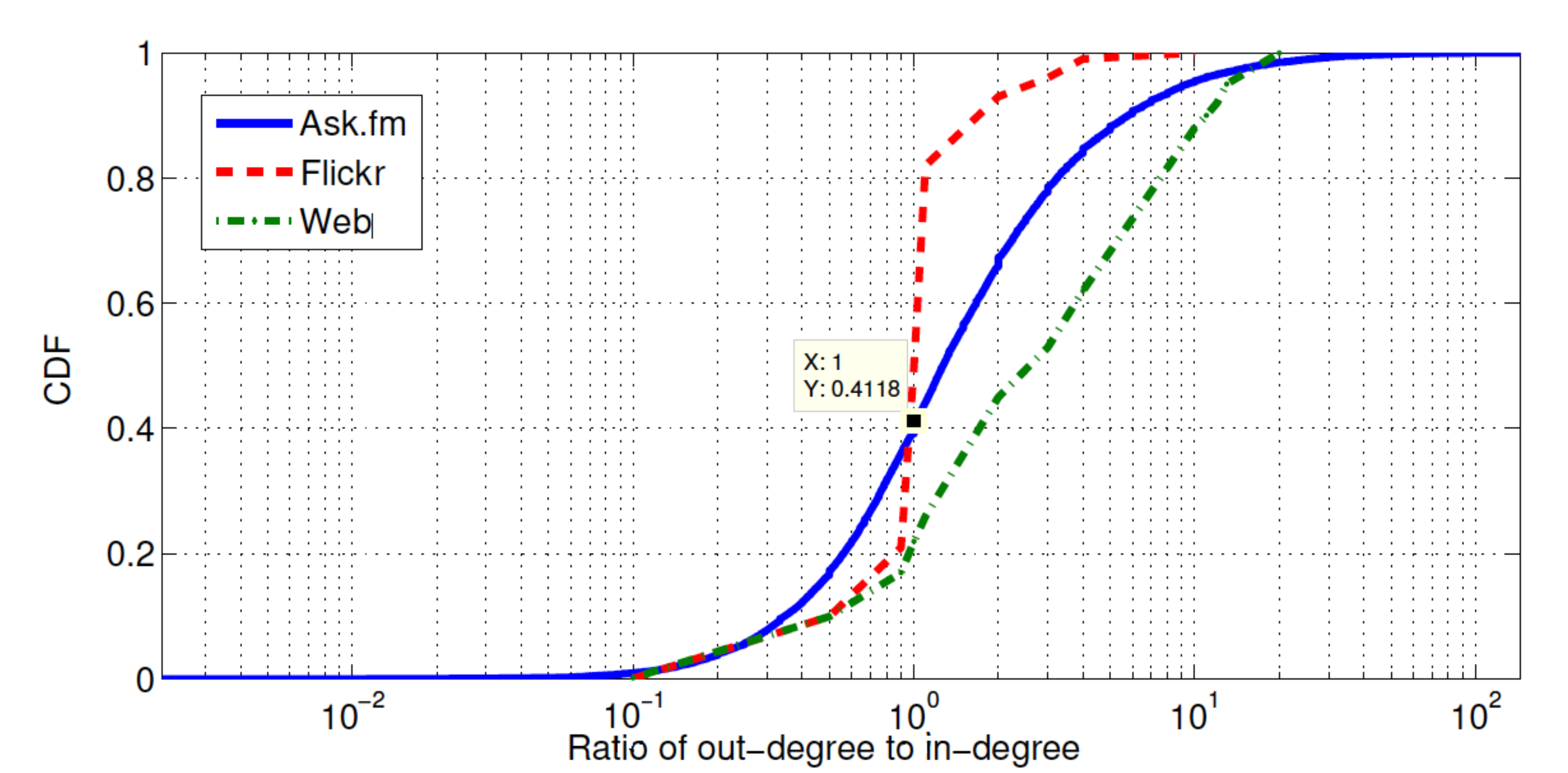}  % use this if you use "pdflatex"
\caption{Cumulative distribution function for ratio of out-degree to in-degree. Measurements for Flicker and Web have been obtained from \cite{Mislove}}
\label{figRatio} % Fig.3
\end{figure} 

Figure \ref{figRatio} shows the CDF for the ratio of out-degree to in-degree. It has been shown in \cite{Mislove} that for social networks more than 50\% of users have in-degree within 20\% of their out-degree. However in the Ask.fm this value is around 16\%. Again we observe that the relationship between in-degree and out-degree in Ask.fm is weaker than what was found for prior social networks and is stronger than the Web. \comment{In the Ask.fm this ratio is around 5\% and in Web is more than 10\% \cite{Mislove}.} \comment{The interaction based behavior have characteristic between Web graph and social graphs and it does not have as high symmetry as social graphs.   }

The mean reciprocity of the interaction graph/matrix $U$ is 28.2\%, which is a low number compared to other social graphs like Yahoo! 360 with reciprocity 84\% and Flickr with reciprocity 68\%, \cite{rcprct}. However Twitter has a more similar structure to Ask.fm (in the sense that there is no friendship concept between users) and has an even lower reciprocity equal to 22.1\% \cite{rcprct}.
Ask.fm's network negative reciprocity is a very low 3.61\%, which shows how much users like each others' negative questions. This gives an insight that users who both have negative posts do not tend to like each others' negative questions. Reciprocity between non-negative questions is a far higher 27.9\%. 

\begin{figure}[htbp!]
\begin{minipage}[b]{0.45\linewidth}
\centering
\includegraphics[width=\textwidth]{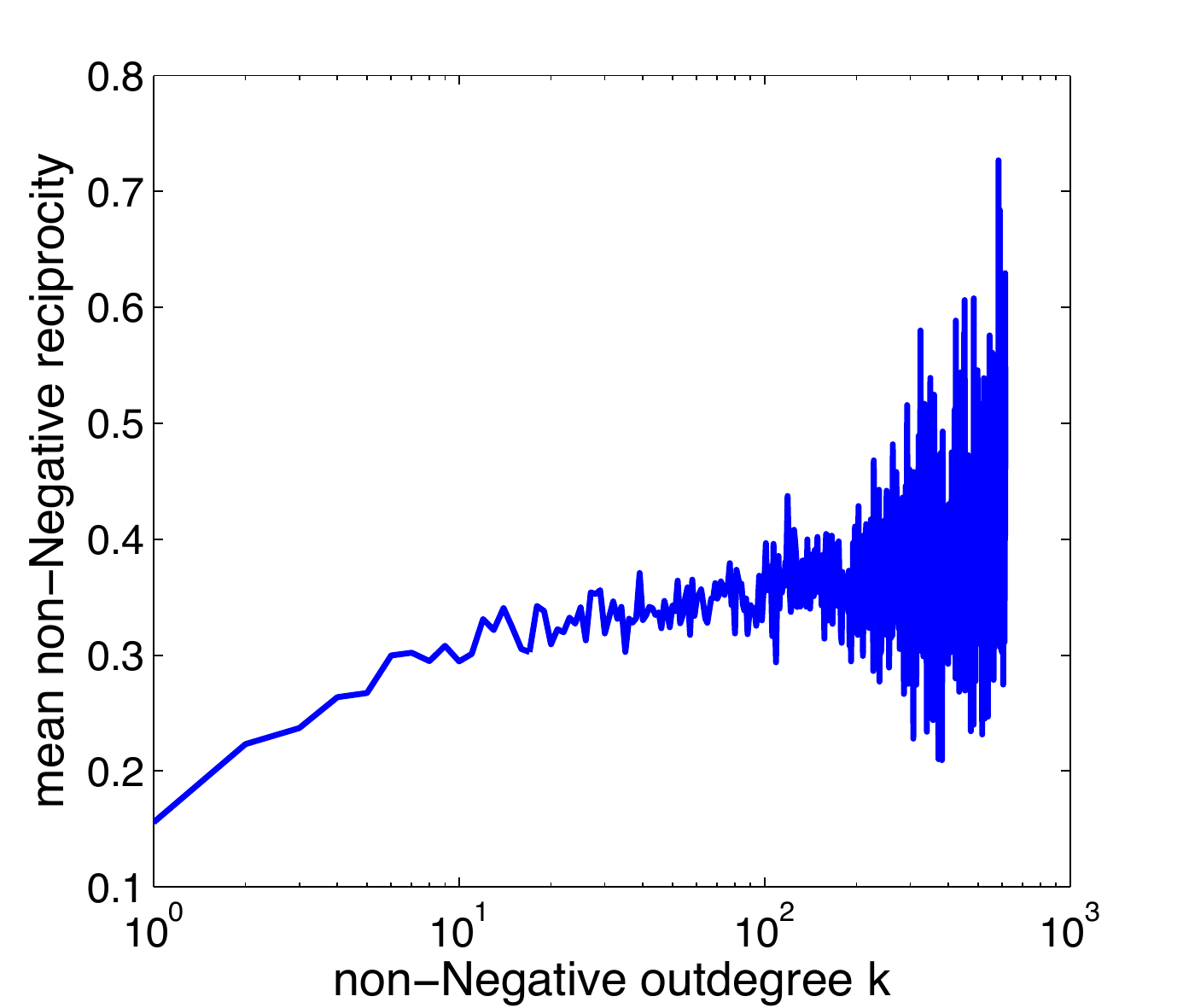}
%\caption{Typical public profile on Ask.fm with questions/comments (usually anonymous) to the profile owner and answers from the profile owner.}
\label{fig:mobile1}
\end{minipage}
\hspace{0.15cm}
\begin{minipage}[b]{0.45\linewidth}
\centering
\includegraphics[width=\textwidth]{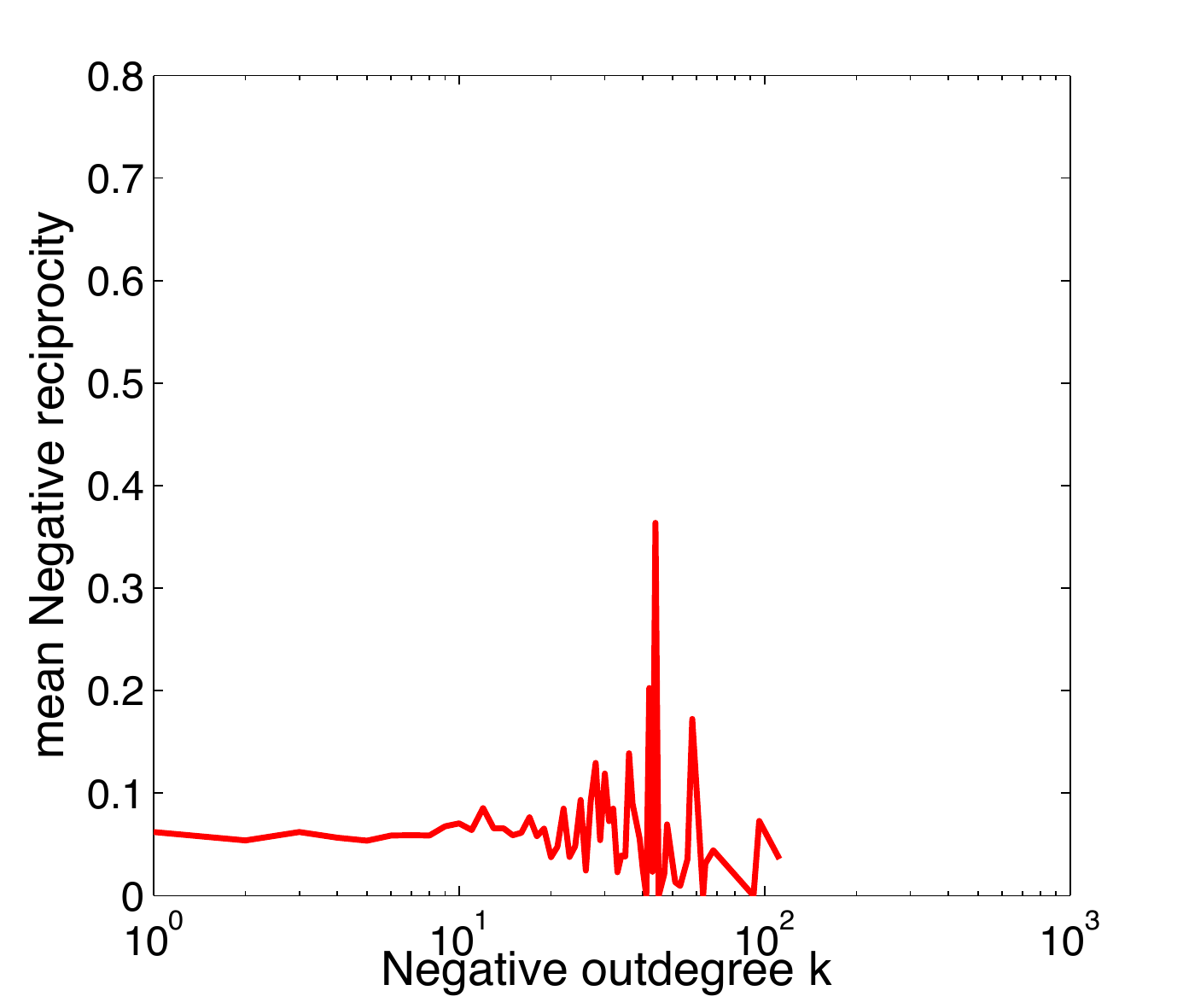}
%\caption{An example of anonymous cyberbullying comments posted on a user's profile in Ask.fm.}
\label{fig:mobile2}
\end{minipage}
\caption{Reciprocity for non-negative and negative interaction graphs/matrices}
\label{fig:recip}
\end{figure}

Figure \ref{fig:recip} shows the mean reciprocity for the two graphs/matrices $U_{neg}$ and $U_{non-neg}$ versus out-degree. It shows that in normal (non-negative) behavior the more active a user is, the more he/she will receive likes from other users or vice versa. On the other hand, the mean reciprocity does not increase with out-degree in the negative matrix. When the negative out-degree is high, this means that a user is active and likes others user's negative questions\comment{(we do not know for sure as support or to bully them)}. However this type of active user receives a low negative in-degree in return. We consider two likely possibilities that explain this result.  First, the user is supportive and popular, liking answered questions often to show support for the answerer of a negative question.  Due to the lack of liking granularity in Ask.fm, this like is recorded as a like of a negative question, even though the intent of the user was to like the answer.  Indeed, we found cases on Ask.fm where a profile owner asked a liker why they liked a negative question, and the liker responded that in fact they were liking the profile owner's response instead.  Such a supportive user would be unlikely to receive many negative questions on their profile.  Alternatively, the user is a bully and unpopular, frequently liking other's negative questions, and not receiving many likes of negative questions in return. \comment{, because they receive few comments at all on their profile.}  Again, we see that negative behavior follows a different pattern than non-negative behavior.
\comment{To relate the preceding graph analysis to cyberbullying, we obtained ground truth for cyberbullying by
labeling a set of 150 profiles as examples of cyberbullying. Users of this labeled data set have a minimum of 1 and a maximum of 8 negative questions. The average number of negative questions is 2.8716.  They have in average 4.3333 positive questions, which implies they have support on their wall beside the negative posts. \comment{ Table \ref{stats2} summarizes these results.  We see compared to Table \ref{stats} that the pattern of negative questions is quite different.}
\comment{
\begin{table}[htbp!]
\centering
\begin{tabular}{|c|c|}
\hline 
Average number of questions per user  &  	 14.8718 \\ \hline
Average number of negative questions per user 	  & 2.8716\\ \hline
Average number of positive questions per user   &    4.3333\\ \hline
Average number of negative words per user 	  & 4.7179\\ \hline
Average number of positive words per user   & 6.0769 \\ \hline
\end{tabular}
\caption{Average number of questions and words per labeled victim of cyberbullying}
\label{stats2}
\end{table} }
}

Next, we explore the relationship between the degree of negativity (and positivity) on a profile and the profile's graph properties.
We compute the average number of negative questions in a set of profiles with negative questions that the profile owner answers to show his/her unhappiness about the questions. For example, as we saw in Figure \ref{figrepeated}, the profile owner says ``you broke me again" in response to repeated negative questions. The average number of negative questions in 150 profiles of this type was 2.87 and therefore we chose a threshold of 3 negative questions to define negative groups. In fact, we segmented the user base into 4 different groups:
 
\begin{enumerate}
\item	Highly Negative: users with at least 3 negative posts and no positive posts (HN)
\item	Highly Positive: users with more than 10 positive posts (HP)
\item	Positive-Negative: users with at least 3 negative posts and more than 4 positive posts (PN)
\item	Others (OTR)
\end{enumerate}

\begin{figure}[htbp!]\centering
\includegraphics[width=0.45\textwidth]{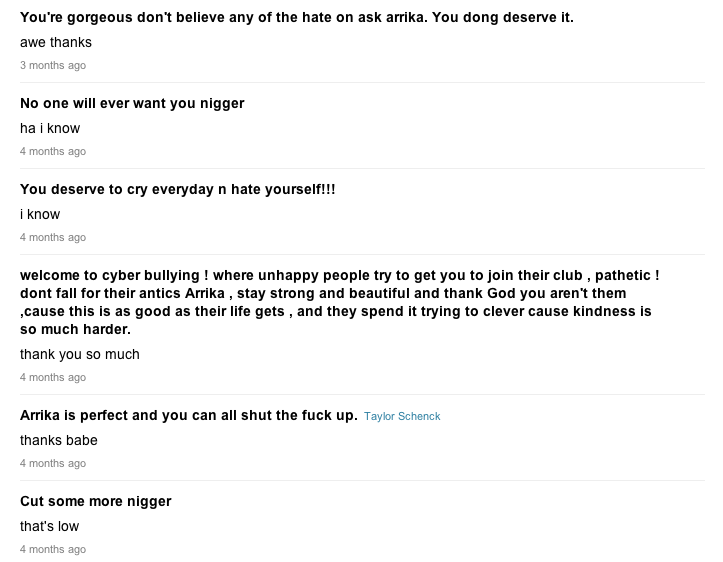}  % use this if you use "pdflatex"
\caption{An example of a profile with both positive and negative comments.}
\label{figRec} % Fig.3
\end{figure}

The definition of these groups helps us identify the properties of the most negative profiles, namely the HN users who have no positive support, while also allowing us to contrast them with the graph properties of other users who have some or a lot of positive words in support.  They were chosen based on our observations that targets of negative behavior could be roughly divided into a group that receives support from bystanders and a group that has been left alone and doesn't receive any support with positive posts.  \comment{We expect groups 1 and 2 to capture most cases of cyberbullying.}  Figure \ref{figRec} shows an example of a mixed profile with both positive and negative comments, representative of the PN group.

\comment{
In fact group 1 and group 2 both are affected by cyberbullying, as they have at least three negative questions and it will be counted as a repeated behavior. Also we are not claiming that group 1 and 3 cover all cyberbullying cases. Group 4 can also have targets of implicit cyberbullying. The goal of separating to these groups is looking closely at users with some similar statistics and evaluating their other features. We observed that victims could be divided in a group who receive support from bystanders and a group who have been left alone and don't receive any support with positive posts, and therefore we get the idea of considering group 1 and 2.
}
\begin{table}[htbp!]\centering
\begin{tabular}{lclclclclclcl} %\hline
                        & HN        & 	HP	     & PN	     &     OTHR   \\ \hline
Negative reciprocity     &  0.131     &   	0     &  0.106    &  0.066  \\
Non-Negative reciprocity	 & 0.210     &  0.379    &  0.246    &  0.339   \\ %\hline
Negative in-degree	     & 6.39     &     0     & 14.0    &  2.16   \\ %\hline
non-Negative in-degree	 & 21.1    & 112  & 49.4    &  70.1	  \\ %\hline
Negative out-degree      & 	3.48    & 2.73    &  3.73    &  3.59    \\ %\hline
non-Negative out-degree	 & 32.8    & 117  & 46.0    & 70.1	   \\ %\hline
Total number of likes	 & 1027    & 5995  & 1765    & 2906	   \\ %\hline
Ratio of likes per answer  &  1.59    &  2.58   &   1.94   &   2.73 \\
\end{tabular}
\caption{Average reciprocity and degree for different groups}
\label{table:keyresults}
\end{table} 

Table \ref{table:keyresults} summarizes some of the key results of our graph analysis.  We measure the average reciprocity of the four different groups, as well as the in-degree and out-degree, based on the interaction graphs $U_{neg}$ and $U_{non-neg}$ calculated earlier.  Note the in-degree has two subcategories, pertaining to negative in-degree and non-negative in-degree, i.e. a user X will have a negative in-degree if another user likes one of the questions posted on X's profile that has a negative word in it, while another user liking a question without a negative word will count towards the non-negative in-degree of X.  Similarly, out-degree has negative and non-negative subcategories.

\emph{From Table \ref{table:keyresults}, a key finding is that HN users are distinguished by having the smallest total (negative plus non-negative) in-degree and the smallest total out-degree.} It shows they are either not popular in terms of being liked or do not tend to have activity in this social network in terms of liking others. That is, the high degree of negativity that these users are subject to is related to less sociable behavior on this social network.  Our results indicate that we could leverage low total in-degree and low total out-degree of Ask.fm's interaction graph to suggest a greater likelihood of determining highly negative profiles of cyberbullying victims on this semi-anonymous social network.

\emph{Another major finding is that as the amount of positive support increases, we find a greater in-degree and greater out-degree of the users, that is users become more social and actively like and are liked more often.}  This is demonstrated first by the PN group, which like the HN group has at least 3 negative posts, but the PN group also has positive support.  We observe that the PN group has increased activity in terms of about twice the amount of total in-degree and total out-degree as the HN group.  Highly positive HP profiles exhibit the highest sociability in terms of actively liking other profiles and being liked by other profiles.  We see that the total in-degree and the total out-degree are both over 100, which is much higher than the second nearest group OTHR, which has a total in-degree and out-degree around 70 each.  This confirms the trend that higher positivity is strongly related to higher sociability on Ask.fm's interaction graph.

An interesting result of this analysis is that we can distinguish the HN group from the PN group not merely by the total in-degree and total out-degree, but also by just the negative in-degree.  We see that the negative in-degree of HN is 6.39, while PN's negative in-degree is 14.0, over twice as large.  
\comment{Perhaps the most striking difference for this group is that the negative in-degree is much lower than the other groups.  There is very little positive support for such users.} %We observe that HN users have a very similar non-negative in-degree to labeled victims of cyberbullying shown in the far right column.  As a result, a key finding is that a low non-negative in-degree may be a useful indicator for detecting cyberbullying victims who are especially vulnerable to abusive behavior.
%The mean number of negative questions for group HN is 3.72 and for group PN is 3.73\comment{, which approximately the same as at the first place we assigned the same limitation on number of negative questions for both of these groups}. On the other hand, the average number of positive questions in the HN and PN groups are respectively 0.76 and 6.52, while each group has on average a total number of questions 14.7 and 15 respectively.  Even though the number of negative questions is similar, the negative in-degree of the PN group is twice as large as the HN group's.  
While the mean number of negative questions is the same in both HN and PN, i.e. about 3, why would there be so many more likes for the same number of negative questions for the PN group?  The explanation for this lies again we believe with the limited granularity of liking in Ask.fm.  In the case of liking a question+answer pair, where the question/comment is negative and the answer is positive, then it would not be clear whether a user is liking a bullier's negative comment or supporting a positive answer from the target of bullying.  The higher volume of likes of negative questions we believe is actually users liking the target's positive response.  This agrees with an examination of a variety of examples, where we saw that the likes were mostly representative of support and a positive sign in this case.

We also observe that HP users have little interaction with negative posts.  That is, their negative out-degree is clearly lower than any of the other three groups.  That means that not only do HP users not have negative questions posted on their profiles, but \emph{HP users spend very little effort liking negative questions on other users' profiles, focusing the vast majority of their effort on liking positive questions.}
\comment{
For the other groups, HP users have the smallest negative out-degree, which shows they are least likely to like negative posts on others profiles.  As expected, HP users have a high non-negative in-degree, but  their non-negative out-degree is much higher than the other users, which indicates that HP users are the most active group in terms of liking other non-negative posts.
%most active group. It means HP users not only do not have negative posts, but also do not like negative posts or do not have that much friends with negative posts.  when we look at negative out-degree, 
For the mixed PN group, perhaps the most distinguishing feature is that the negative in-degree of the group PN is more than twice that of the next closest group HN.
}
Looking at negative reciprocity, it is clear that HP has negative reciprocity 0 as it does not have any negative in-degree. In the group OTHR it seemed that there existed either a set of users with 1 or 2 negative questions without any limitation on the positive posts, or a set of users with more than 2 negative questions and less than 10 positive questions. However, the average number of negative questions is 0.774, which shows we have mostly users belonging to the first set. We see that the reciprocity on negative questions in this group is lower than the HN and PN groups. The reason is either because they have few negative questions compared to the HN and HP groups (which in average have less), or they have less support (they have in average 4.76 positive questions compared to 6.53 positive questions for the HP group). 

\begin{figure}[htbp!]\centering
\includegraphics[width=0.45\textwidth]{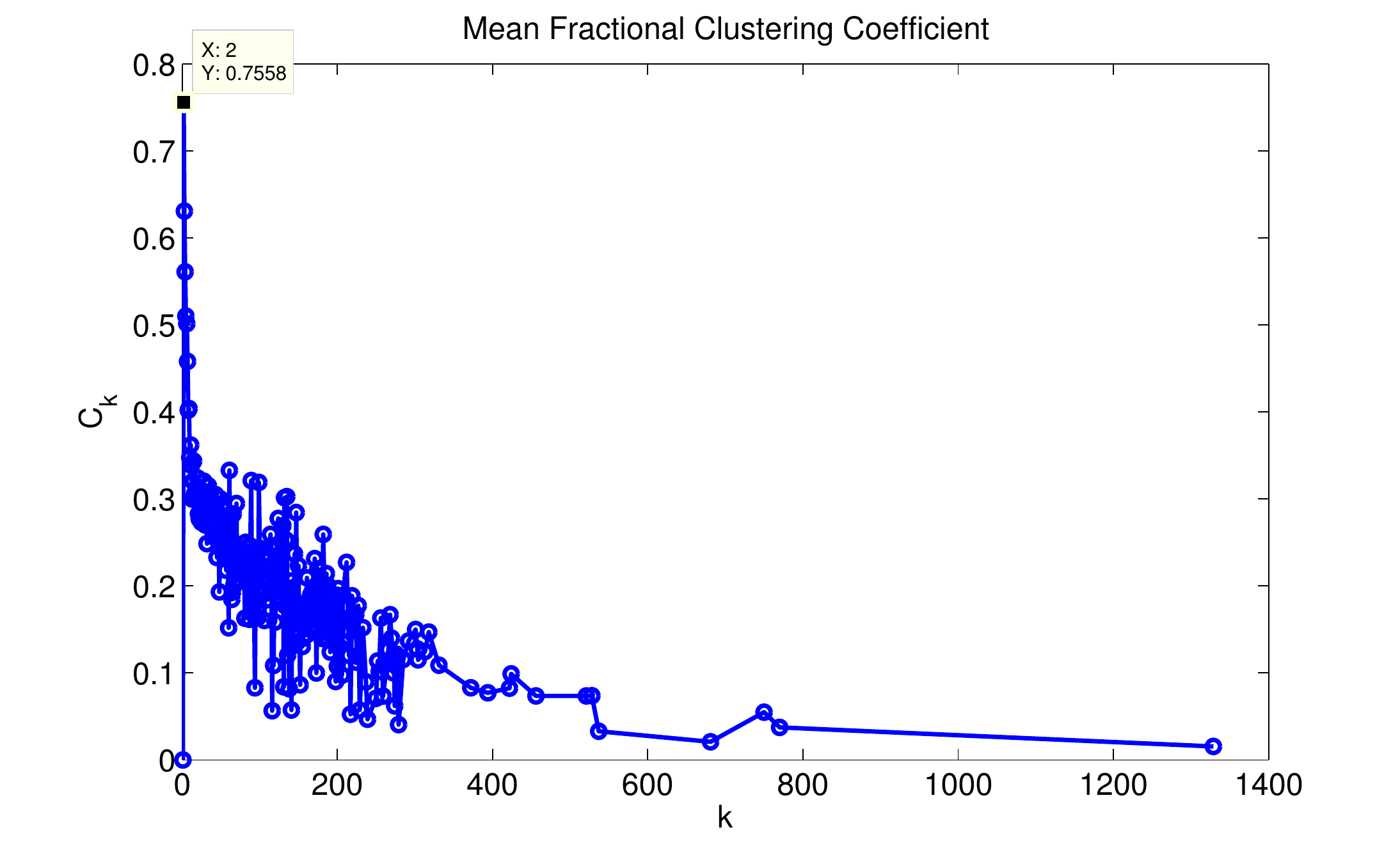}  % use this if you use "pdflatex"
\caption{Mean local clustering coefficient versus degree}
\label{LCC} % Fig.3
\end{figure}

In order to calculate the local clustering coefficient for each node we first turned our network into a simple graph. It means either node $i$ has liked node $j$'s question, or node $j$ has liked node $i$'s question, regardless of the number of likes.  We set $U_{ij} = 1$ and $U_{ji} = 1$, otherwise $U_{ij} = U_{ji} = 0$. We make this assumption that either user $i$ receives a like from user $j$ or posts a like on his/her profile, they are in the general category of having some familiarity or ``connectedness''. The expected clustering coefficient of the network is 0.11 \comment{using formula 7.41} and the averaged local clustering coefficient 0.356\comment{using formula 7.44 of the book}, as defined in  \cite{netbook}. Comparing with numbers reported by \cite{data}, shown in Table \ref{LLC_web}, the clustering coefficient of Ask.fm is pretty small. In Figure \ref{LCC} the local clustering as a function of degree has been depicted. As we expect, the local clustering coefficient decreases when the degree increases in social networks. 
\begin{table*}[htbp!]\centering
\centering
\begin{tabular}{|c|c|c|c|c|c|c|c|c|c|}
\hline 
social media           & Ask.fm     & Facebook    & Twitter & 	Gplus    & Flickr & Orkut & You Tube \\ \hline
Clustering Coefficient & 	0.356 & 	0.606	 & 0.565  & 	0.490	 & 0.313   & 0.171     & 0.136 \\ \hline 
%Clustering Coefficient & 	0.1095 & 	0.2647 & 	0.06415 & 	0.6552 & 	0.01414\\ \hline 
\end{tabular}
\caption{Local clustering coefficient for some social networks. Clustering coefficient values for Flickr, Orkut and You Tube are from paper \cite{Mislove} and values for Facebook, Twitter, Gplus are from the website \cite{data}. }
\label{LLC_web}
\end{table*}
Looking at the clustering coefficient of each group in Table \ref{tab:weblog-data}, we observe that among 4 defined groups, HN has the highest mean local clustering coefficient despite having the lowest degree, while the group HP with highest degree has the lowest mean local clustering. The implication is that users of the HN group probably have a few people that know each other. However, users of the HP group are very social and know many people, and therefore the proportion of their friends that know each other is small. In general, we can see each group that has higher degree has lower mean local clustering. \comment{ However, for the labeled victims of cyberbullying, we can see the mean local clustering is smaller than HN, although its out degree is smaller than the HN group.}

\begin{table}[htbp!]
\centering
\begin{tabular}{|c|c|c|c|c|}
\hline 
	HN & 	HP	 & PN & 	OTHR	 \\ \hline
 	0.499    &  	0.237   &   	0.380    &  	0.357	\\ \hline

\end{tabular}
\caption{Local clustering coefficient for defined group}
\label{tab:weblog-data}
\end{table}
%\comment{
%}
\comment{
\subsection{Temporal Behavior}
Looking back at the profiles after 3 months,  14.4\% profiles were deactivated which compare to before it has increased 5\%.

out 7.87\% of active users deactivated their account during past 3 months. Also 10\% of users with more than 3 negative questions deactivated their profiles

\begin{table*}[htb]
\centering
\begin{tabular}{|c|c|c|c|}
\hline 
  & all users & users who deactivated their accounts &  users who their accounts have been suspended\\ \hline
Average total number of lines per user         & 3.0067e+03 & 3.8258e+03 & 8.0421e+03\\ \hline
Average total number of answers per user       & 1.1237e+03 & 1.4335e+03 & 1.4297e+03\\ \hline
Average number of answers per user              &  	 14.34	 &  	 14.6	& 14.7 \\ \hline
Average number of negative questions per user   & 0.93	  & 0.88 & 1.44\\ \hline
Average number of positive questions per user    &    4.83  &    5.16 & 5.2\\ \hline
Average number of negative words per user & 1.25	& 1.19 & 2.1\\ \hline
Average number of positive words per user & 6.1  & 6.52 & 6.7\\ \hline
\end{tabular}
\caption{Average number of questions and words per user for users who deactivated their profiles}
\label{stats}
\end{table*}

 7\% of HN, 9.8\% NP, 8\% of HP and 7\% of normal users deactivate their accounts. Among users whose account has been suspended, 0.4\%   10.17\%   23.4\%   65.92\% , however users whose accounts was deactivated         0.39    13.47\%    10.69\%    75.45\% 
 }

%% file: hrisk.tex
% !TEX root = main.tex
\section{Users with Cutting Behavior}
\begin{figure}[htbp!]\centering
\includegraphics[width=0.45\textwidth]{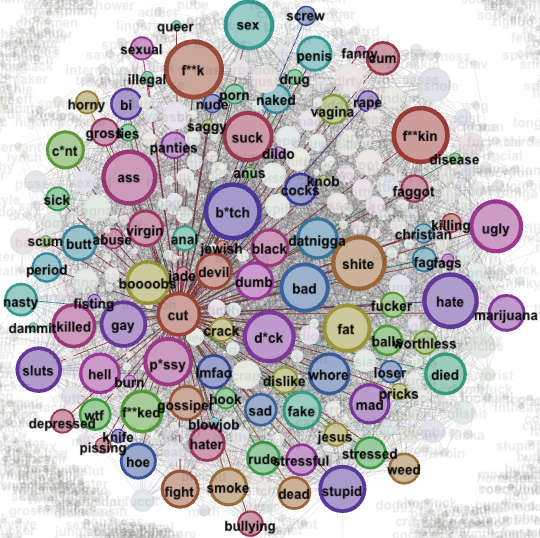}  % use this if you use "pdflatex"
\caption{Word usage with the word ``cut" in Ask.fm. The size of the circle indicates Eigenvector Centrality score}
%, visualized using Gephi \cite{Gephi}.}
\label{cut1} % Fig.3
\end{figure}

One of the most disturbing behaviors that we encountered was the problem of ``cutting" (slicing one's wrists).  Looking at the profiles of these cases and studying their answers, it seemed that the profile owner exhibited weak confidence and depression problems, sometimes admitting to earlier failed attempts at suicide. In this section we first found 150 profiles for which their owners have explicitly expressed the experience of ``cutting" behavior and label them.  A human labeled the profile as ``cutting" behavior if and only if the profile owner has expressed explicitly in his/her answers that he/she has had such an experience. We observe in Figure \ref{cut1} that among the words that the word ``cut" has been connected to are the words ``depress", ``stressful", ``sad", and ``suicide" . This association could be used to detect these type of users. 

\begin{figure}[htbp!]\centering
\includegraphics[width=0.38\textwidth]{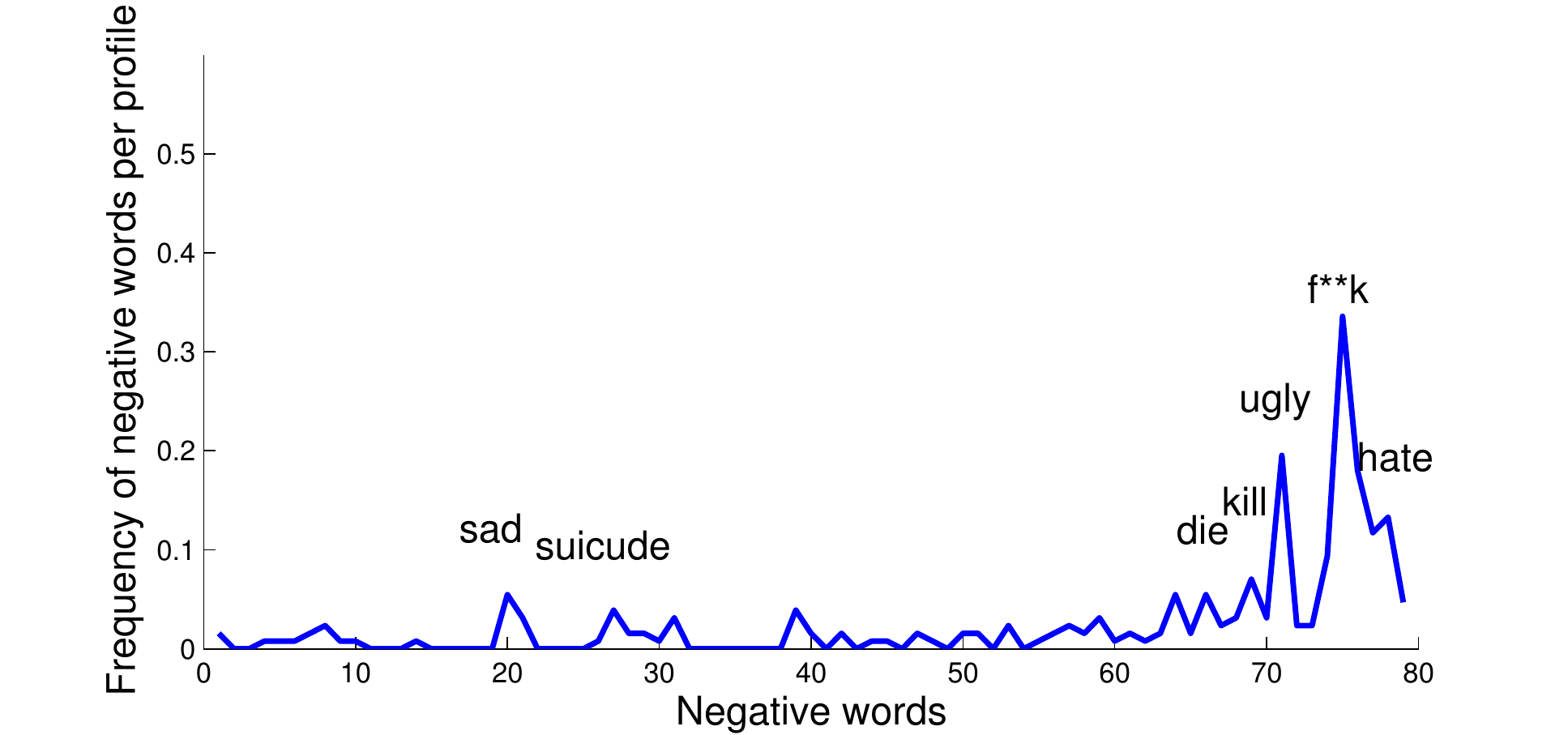}  % use this if you use "pdflatex"
\caption{Frequency of negative words seen with word "cut".}
\label{ct} % Fig.3
\end{figure}

The frequency of negative words used with ``cut" has been shown in Figure \ref{ct}. We can see these user profiles have two peaks at words ``ugly" and ``f**k".

\begin{table}[htbp!]
\centering
\begin{tabular}{|c|c|}
\hline 
Average number of questions per user  &  	 15	\\ \hline
Average number of negative questions per user 	  & 2.31\\ \hline
Average number of positive questions per user   &    5.29\\ \hline
Average number of negative words per user 	  & 3.38\\ \hline
Average number of positive words per user   & 6.65\\ \hline
\end{tabular}
\caption{Average number of questions and words per user with ``cutting" label}
\label{stats_cut}
\end{table} 

Table  \ref{stats_cut} illustrates the statistics of the labeled users for ``cutting" behavior. We found that such profiles have positive questions close to PN profiles however the average number of negative posts is less by a factor 0.65 compared to the PN and HN groups. This suggests surprisingly but encouragingly that these profiles do receive more support and less negative posts compared to general HN users. 

Table \ref{highrisk} shows the statistics of the average reciprocity, in-degree and out-degree for cutting victims.  \emph{Though we originally expected this group to exhibit behavior similar to the HN group, we found instead that this cutting group appears to exhibit collectively a different behavior than the HN, HP, PN, and OTHR groups measured in Table \ref{table:keyresults}}.  The in-degree is 1.5 times more than PN's in-degree and the out-degree is also more by a factor of 1.4. In fact total in-degree and out-degree is more similar to the group OTHR, though there is a marked difference in negative in-degree compared to OTHR. This group has an average number of negative and positive questions most closely related to PN profiles. However, it receives more likes (higher in-degree), and exhibits more activity (higher out-degree) compared to PN groups. This deserves further investigation to explain the reasons behind the differences.

\begin{table}[htbp!]\centering
\begin{tabular}{lclclclclclcl} %\hline
                        & 	users with ``cutting" label \\ \hline
Negative reciprocity     &  	 0.146  \\
Non-Negative reciprocity	 &   0.248\\ %\hline
Negative in-degree	     & 	17.9 \\ %\hline
non-Negative in-degree	 &  78.7 \\ %\hline
Negative out-degree      & 	 4.42 \\ %\hline
non-Negative out-degree	 &  65.0 \\ %\hline
\end{tabular}
\caption{Average reciprocity and degree for users with ``cutting" label}
\label{highrisk}
\end{table}   
\comment{Looking back at this profiles after 3 months, from 150 profiles, 16 were deactivated and 4 profiles have been suspended. }

%% file: conclusion.tex
% !TEX root = main.tex
\section{Conclusions and Future Work}

As far as we are aware, this paper is the first to present a detailed analysis of user behavior in the Ask.fm social network.  We analyzed nearly 30K profiles of Ask.fm users using interaction graphs, word graphs, and frequency distributions, and characterized key properties such as reciprocity, clustering coefficient, and the influence of negativity on in-degree and out-degree.  Some of the key findings of the work are that 
(1) When people have highly negative profiles without any positive support, they also have the lowest activities in terms of both in-degree and out-degree, that is they are the least sociable.
(2) As the amount of positive support increases, we find a greater in-degree and greater out-degree of the users, that is users become more social and actively like and are liked more often.  For example, users with negative profiles that also receive positive support have higher in-degree and out-degree than owners of highly negative profiles that lack positive support, that is they are more sociable.  When people have highly positive profiles, then they also have the highest in-degree and out-degree, which shows that they socialize the most on this social network.  This suggests that we may be able to use the interaction graph's in-degree and out-degree as an indicator of the extent of negativity or positivity on a given profile.  (3)  The negative in-degree and negative out-degree do not exhibit similar behavior, unlike the similarity of in-degree and out-degree found in other social networks.  (4)  Finally, our analysis of cutting behavior on Ask.fm reveals that such profiles have surprisingly high positive support, and exhibit a different signature than the other group segments studied.

%the profiles of victims of negative user behavior are strongly related to profiles that have repeated (3) negative questions/comments, and that vulnerable victims of cyberbullying can be identified by a lack of positive support from other users.  

\comment{In addition, we have characterized different types of cyberbullying behavior, including religious, appearance-based, LGBT, and racial cyberbullying, and found that many of them can be distinguished by the different shapes or heights of distributions of negative words associated with a core word representing that type of distribution.  We have also identified that users associated with ``cutting" exhibits have a different pattern than general cyberbullying.}

\comment{ partial picture: not all outgoing links, only top 15 likes, can't find followers nor who you're following, ...}

\comment{
 The goal in this paper was looking for patterns of negative behavior. We can see pattern of users who receive negative questions. However we could not find any pattern for people who post them. First, because all the questions have been posted anonymously, therefore we could not find who has posted the bullying post. Secondly, at the starting point we had the assumption that whoever likes a negative post, means that is confirming that negative comments about the profile owner. However, after looking at the relation between number of positive posts and negative in-degree, we believe that these likes are not coming from bullies, instead come from the bystanders, mostly to support the answer of profile owner. 
}

Our initial analysis of cyberbullying on Ask.fm has uncovered a plethora of future research directions.  We hope to design classifiers to detect cyberbullying and evaluate their accuracy and false positives/negatives over the general user population on Ask.fm.  Another way to improve detecting victims is looking at the answers in addition to the question. Here by not inspecting the content of the answers, we are potentially missing useful information. Answers give us further insight on when a behavior received from other users starts to disturb the profile owner. \comment{Following temporal information is another important sign of the presence of cyberbullying.  However, such analysis adds high complexity. The frequency of receiving negative and non-negative posts in a period of time gives us a helpful feature in victim detection. }  Also, we have only looked at the top 15 liked questions. We can extend this to all questions, and also investigate the role of the most recently posted questions on a user's page.  \comment{We would like to develop more detailed models for each type of cyberbullying, fusing together the patterns around different words for each type to develop a model for each type, and then match a given profile to determine the best type.}\comment{We also plan to investigate mixed types, where more than one type of cyberbullying may be going on simultaneously.}\comment{We further intend to explore the impact of the severity of negativity in a question on cyberbullying.} We further intend to conduct a more extensive sampling of Ask.fm, obtaining a larger set of profile data from this social network.  We would like to investigate in more detail the characteristics of high risk cutting victims to ease their identification.  In order to determine the effect of anonymity on the degree of negativity in user behaviors, we intend to compare the semi-anonymous social network Ask.fm with non-anonymous social networks.

%two negatives in a single question - severity, building a word graph based on a word in question.

%consider non-negative words as well

\comment{	
	Looking at the posts, we observed that a lot of users start deactivating their accounts, especially due to the prevalence of bullying behavior in Ask.fm. We can have this in mind in order to determine the generative model of vanishing in this social network.  This will require gathering data over the long term.
	} 
	
\section*{Acknowledgment}
We wish to acknowledge financial support for this research from the US National Science Foundation (NSF) through grant CNS 1162614.